\documentclass[a4paper,noarxiv,onecolumn,accepted=2022-06-02]{quantumarticle}

\usepackage{amsmath,amssymb}
\usepackage{amsthm}
\usepackage{physics}
\usepackage{graphicx}
\usepackage{hyperref}
\usepackage{xcolor}
\usepackage{geometry}
\usepackage{xcolor}
\theoremstyle{plain}
\newtheorem{theorem}{Theorem}

\newtheorem{prop}[theorem]{Proposition}
\newtheorem{coro}[theorem]{Corollary}

\usepackage[capitalize]{cleveref}
\usepackage{tabularx}
\usepackage{algorithm}
\usepackage{algpseudocode}

\theoremstyle{definition}

\theoremstyle{remark}
\newtheorem{remark}{Remark}

\newcommand{\wh}[1]{\widehat{#1}}

\newcommand{\veps}{\varepsilon}
\newcommand{\CO}[1]{\mathcal{O}\left(#1\right) }

\newcommand{\dt}{\Delta t}
\newcommand{\dx}{\Delta x}
\newcommand{\px}{\partial_x}

\newcommand{\ef}[1]{\exp\left( #1 \right)}

\definecolor{orange}{rgb}{1,0.5,0}

\usepackage{lipsum}
\usepackage{mdframed}
\usepackage{booktabs}
\usepackage{longtable}

\begin{document}

\title{Quantum simulation in the semi-classical regime}

\author{Shi Jin} 
\affiliation{School of Mathematical Sciences, Institute of Natural Sciences, MOE-LSEC and SHL-MAC, Shanghai Jiao Tong University, Shanghai, China} 
\email{shijin-m@sjtu.edu.cn}

\author{Xiantao Li}
\affiliation{Department of Mathematics, Pennsylvania State University, University Park, PA 16802, USA}
\email{ Xiantao.Li@psu.edu}

\author{Nana Liu} 
\affiliation{Institute of Natural Sciences, University of Michigan-Shanghai Jiao Tong University Joint Institute, MOE-LSEC, Shanghai Jiao Tong University, Shanghai, China}
\email{Nana.Liu@quantumlah.org}


\begin{abstract}
{
Solving the time-dependent Schr\"odinger equation is an important application area for quantum algorithms. We consider Schr\"odinger's equation in the semi-classical regime where $\hbar \ll 1$ and the solutions exhibit strong multiple-scale behavior due to a small parameter, in the sense that the dynamics of the quantum states and the induced observables can occur on different spatial and temporal scales.   Such a Schr\"odinger equation finds applications in many fundamental problems in quantum chemistry,  including those from Born-Oppenheimer molecular dynamics and Ehrenfest dynamics. However, the presence of the small parameter $\hbar$ in these applications compromises both the spatial resolution and the time integration accuracy. Therefore, to assess a Hamiltonian simulation algorithm in this regime, it is important to quantify the  complexity   in terms of both $\hbar$ and an error tolerance $\veps$.    This paper considers quantum analogues of pseudo-spectral (PS) methods on classical computers.  Estimates on the gate counts in terms of $\hbar$ and the precision $\veps$ are obtained. It is found that the number of required qubits, $m$, scales only logarithmically with respect to $\hbar$. When the solution has bounded derivatives up to order $\ell$, the symmetric Trottering  method has gate complexity  $\CO{ (\veps \hbar)^{-\frac12} \mathrm{polylog}(\veps^{-\frac{3}{2\ell}} \hbar^{-1-\frac{1}{2\ell}})},$ provided that the diagonal unitary operators in the pseudo-spectral methods can be implemented with $\mathrm{poly}(m)$ operations.  When physical observables are the desired outcomes, however,  the step size in the time integration can be chosen  {\it independently} of $\hbar$.  The gate complexity in this case is reduced to  $\CO{\veps^{-\frac12} \mathrm{polylog}( \veps^{-\frac3{2\ell}} \hbar^{-1} )},$ with $\ell$ again indicating the smoothness of the solution. 
}
\end{abstract}

\maketitle

\section{Introduction}

We consider the time-dependent Schr\"odinger equation (TDSE) in the semi-classical regime
\begin{equation}\label{eq: schr}
\begin{aligned}
&  i  \partial_t\ket{\psi} = H \ket{\psi}, \quad H:=- \frac{\hbar \nabla^2}{2} +  \hbar^{-1} V(x), \\
&  \ket{\psi(0)} = \ket{\psi_0}.
\end{aligned}
\end{equation}
In \cref{eq: schr},  the $-\nabla^2/2$ term corresponds to the kinetic energy, while $V(x)$ is the external potential.  Compared to standard TDSE in quantum chemistry with $m=1$, $\hbar$ in \cref{eq: schr} plays the same role as the Planck constant. The semi-classical regime refers to the case when  $\hbar \ll 1.$  This is one important example of multiscale quantum dynamics, where the small parameter leads to fine scale dynamics, but one is interested in the behavior on a much large scale.  
%We are interested in the dynamics on a scale much larger than that represented by the small parameter, i.e., $\hbar \ll 1.$ 

Eq. \eqref{eq: schr} arises in various important applications \cite{jin2011mathematical}. Perhaps the most notable example is the Bohn-Oppenheimer approximation \cite{BO27, Tul90, Zen32}, where  Eq. \eqref{eq: schr} describes the dynamics of nuclei and $\hbar^2$ is the ratio between the electron and nuclei mass. {A small value of $\hbar$ enables the Bohn-Oppenheimer approximation, which separates the nucleonic dynamics from that of electrons and also allows the classical approximation to the nucleonic Schr\"odinger equation}. {It enters the effective nuclear Schr\"odinger equation in the form of  \eqref{eq: schr}. } It also arises in Ehrenfest dynamics, where \eqref{eq: schr} is the electronic Schr\"{o}dinger equation, which is
coupled with a classical dynamics for nuclei through $V$, and $\hbar^2$ again is the ratio between the electron and nuclei mass \cite{BNS96, Dru99, Hin73}.  One can also find related  applications in geometric optics, where one is interested in  the WKB solution 
\begin{equation}\label{wkb}
\ket{\psi_0}= A_0(x) \exp \left( i \frac{S_0(x)}{\hbar} \right).
\end{equation}
Eq. \eqref{eq: schr} leads to the well-known  eikonal  and transport equations  \cite{jin2011mathematical}. In the WKB form, $A_0$ and $S_0$ are often referred to as the amplitude and phase function, respectively. 
In light of these applications, we will regard $\hbar$ simply as a small parameter, rather than just the Planck constant in time-dependent Schr\"odinger equations.

A typical observation from the solution of Eq. \eqref{eq: schr} is that rapid oscillations exist with wavelength comparable to $\hbar$. This poses a great challenge for direct numerical simulations since one needs to numerically resolve the fine scale dynamics. This means that the number of grid points and time steps must be both inversely proportional to $\hbar.$ One popular classical algorithm for \eqref{eq: schr} is the Fourier pseudo-spectral method \cite{bao2002time}. Here we briefly demonstrate the procedure for the one-dimensional case.  Assume that the spatial domain is $[0,L]$, with periodic boundary conditions applied at the boundary $x=0$ and $x=L$. The wave function
is then discretized in space using a uniform grid size $\dx=L/M.$ The grid points are given by,
\[ x_j= j\dx, \; j=0,1,2,\cdots, M.\]  
In addition, the time steps will be denoted by $t_n = n \dt,$ with $\dt$ being the step size. We also write $\ket{\psi^n}= \ket{\psi(\cdot,t_n)}.$

From the time step $t_n$ to $t_{n+1}$, one can use the Strang splitting,
\begin{equation}\label{eq: splitting-sp}
  \ket{ \psi^{n+1} } \approx  U_K\left(\frac{\dt}{2}\right) U_V\big(\dt\big) U_K\left(\frac{\dt}{2}\right)  \ket{\psi^n}. 
\end{equation}
Here
$ U_V(\dt):= \ef{-i \frac{\dt}{\hbar} V}$ is understood as a diagonal operator, i.e., applying $ \ef{-i \frac{\dt}{\hbar} V(x_j)}$ to $\ket{\psi^n}$ at $x_j$. Meanwhile, we have defined 
\begin{equation}
    U_K(\dt)= \ef{i\frac{\dt \hbar   \nabla^2}{2}}.
\end{equation}

Since the kinetic energy part $K$ can be diagonalized by a Fourier transform, the operator  $  U_K$ can be efficiently implemented by a finite Fourier transform:  
\begin{equation}\label{eq: fft}
  \braket{j}{ \psi} = \frac1{\sqrt{M}} \sum_{k=0}^{M-1}  \exp(i\frac{ 2 \pi j k}M )     \braket{k}{\psi}.
\end{equation}
On classical computers, this is implemented using the Fast Fourier Transform  (FFT) with complexity $M\log M$. In order to look at the algorithm from a quantum computing perspective, we have used $\ket{j}$ as the computational basis. 

With the Fourier transform \eqref{eq: fft}, an application of $U_K$ will result in
\begin{equation}\label{eq: kin}
  \bra{j} U_K(\dt) \ket{\psi} = \frac1{\sqrt{M}} \sum_{k=0}^{M-1}  \exp( -i  \frac{\dt\hbar}{2} \mu_k^2 )   \exp(i \frac{2 \pi j k}M )     \braket{k}{\psi},
\end{equation}
where
\begin{equation}\label{eq: mu}
    \mu_k = \dfrac{2\pi }{L}(k-\frac{M}2), \quad k=0, 1, \cdots, M-1.
\end{equation}
Therefore, the implementation involves one application of a FFT to obtain $\ket{\psi}$, followed by a multiplication of a diagonal unitary operator ( $\exp( -i  \frac{\dt\hbar}{2} \mu_k^2 )$), and
an inverse FFT. Each step only involves $\mathcal{O}(M\log M)$ operations. 

The algorithm outlined above can be viewed as a Trotter splitting. Although Trotter type splitting methods, in the context of quantum algorithms, have been widely studied \cite{childs2021theory},   compared to the usual time-dependent Schr\"odinger equation, the model \eqref{eq: schr} presents several new challenges. 
\begin{itemize}
    \item The initial condition, as well as the solution at later times,  involve $\hbar$, which is often manifested in the fast oscillations in the wave function. As a result, a numerical analysis of the truncation error will contain powers of $\hbar^{-1}$.
    \item The two operators in $H$ from \eqref{eq: schr} are of different magnitude: the kinetic energy term  is of $\mathcal{O}(\hbar)$ while the potential energy is of the order $\mathcal{O}(\hbar^{-1})$. This complicates the analysis of the Trotter error; yet an $\hbar$-dependent error analysis is important in many applications such as the Born-Oppenheimer approximation, the Ehrenfest dynamics, or in general, quantum dynamics in the semi-classical regime.
    \item In practice, often of interest is the dynamics of physical  observables, rather than the wave function itself. Although an error bound for the wave function can be carried over to observables, such a bound can be an  overestimation, especially for a multiscale problem like \cref{eq: schr}, since the fast oscillations may be averaged out in observables. Therefore, error analysis that specifically targets observables are needed to derive more accurate  complexity estimates. 
\end{itemize}

Overall the presence of the small parameter $\hbar$ compromises both the spatial resolution and the time integration accuracy. Therefore, to assess a Hamiltonian simulation algorithm in this regime, it is important to quantify the  complexity   in terms of both $\hbar$ and an error tolerance $\veps$. To the best of our knowledge, this has not been studied in the literature. 

\noindent{\bf Related works.} Childs et al. \cite{childs2021theory} considered general Trotter splitting methods, and provided a gate complexity for a  $p-$th order Trotter splitting algorithm. They also examined the case when the kinetic energy and potential energy terms can be represented separately in the Fourier and real spaces, respectively. They find that for physical observables, the gate number can be significantly reduced. { A common practice in such analysis of Trotter error is to bound the error using the operator norm. However, in the real space representation, the kinetic energy operator is unbounded. Even with a spatial discretization, the norm typically scales like $\Delta x^{-2}$ with $\Delta x$ being the grid size. What further complicates the error analysis in the semiclassical regime is the fact that the wave functions are highly oscillatory, with derivatives $\partial_x^k \ket{\psi}=\CO{\hbar^{-k}}$. Therefore, the commutators alone can not represent the magnitude of the error.  }

\medskip 
{
\noindent{\bf Our contributions.} We consider quantum algorithms for the  dynamics in the semiclassical regime \eqref{eq: schr}. 
A natural choice is the pseudo-spectral method, in conjunction with the operator splitting schemes \cite{bao2002time}, as discussed above. The operator splitting schemes alternate between the kinetic and potential energy terms, each of which can be reduced to diagonal unitary operators. Therefore, they can be viewed as Trotter splitting methods. Classical algorithms  typically require $\CO{\hbar^{-d}}$ operations for each step of the time integration. Here $d$ is the dimension and the dependence on $\hbar$ is owing to the fact that the grid size has to resolve the small parameter $\hbar$. In contrast, in quantum computers, this can be efficiently represented by $m$ qubits, and we show that the dependence on  $\hbar$ is only logarithmic.  Provided that the diagonal unitary operators in the pseudo-spectral methods can be implemented with $\mathrm{poly}(m)$ operations, this only leads to  $\CO{d\mathrm{polylog}(\hbar^{-1} ) }$ operations. The dependence on the precision $\veps$, the domain size $L$, and the time duration $t$, will all be included in the estimate.

Using the recent rigorous error estimates  \cite{golse2021convergence,lasser2020computing}, we estimate the gate complexity in order to  reach a given precision $\veps$. For the commonly used symmetric Trottering splitting, we show that the gate complexity is
$\CO{ (\veps \hbar)^{-\frac12} \mathrm{polylog}(\veps^{-\frac{3}{2\ell}} \hbar^{-1-\frac{1}{2\ell}})},$ 
 under the same assumption on the implementation of the diagonal unitary operators and the solution has certain degree of smoothness indicated by the integer $\ell$. In addition, when the solution has better smoothness, in that certain $\hbar$-dependent Sobolev norm of the solution $\| \psi \|_{H_\hbar^q}$, is finite, then gate number has a more relaxed dependence on $\veps$. The better smoothness also allows higher order Trotter splitting algorithms, e.g., the fourth order method by Yoshida \cite{yoshida1990construction}.  In this case, the gate complexity is close to $\CO{ p (\veps \hbar)^{-1/p} \mathrm{polylog}( \hbar^{-1})}.$

A remarkable finding in \cite{bao2002time} is that when observables, rather than the wave functions themselves, are of interest, the time integration error is not as sensitive to the small parameter $\hbar$.  In fact, the step size can be chosen {\it independent} of $\hbar$. This has been numerically verified  in numerous works and   rigorously proved in recent papers  \cite{golse2021convergence,lasser2020computing}. Clearly a larger step size will reduce the overall complexity. We show that for the symmetric splitting method, the gate complexity is reduced to $\CO{\veps^{-\frac12} \mathrm{polylog}( \veps^{-\frac3{2\ell}} \hbar^{-1} )},$ which exhibits a much less dependence on $\hbar$.
Overall, the current work highlights an important direction in Hamiltonian simulations, i.e., quantum dynamics that exhibit multiple physical scales. The development of efficient quantum algorithms must take into account the asymptotic behavior, e.g., when a small parameter approaches zero. 

}

\section{Quantum Algorithms for the  Schr\"odinger equation}

As alluded to in the introduction, the primary challenge in solving \eqref{eq: schr} is due to the small parameter $\hbar$: Solutions typically involve oscillations with wavelength comparable to $\hbar.$ As a result, the grid size $\dx$ has to be chosen small enough to resolve the fine scale, which leads to an overwhelmingly large problem size, especially for high-dimensional cases. Quantum algorithms have the potential to deal with exponentially large dimensions. As an example, for a quantum computer with $m$ qubits, one can choose $M=2^m$ and represent the wave function $\ket{\psi}$ using the computational basis $\ket{j}$. Therefore, the number of qubits depends only logarithmically on the grid size, and linearly on the spatial dimension. Another important observation  is that the FFT \eqref{eq: fft} can be implemented using a quantum circuit. The gate complexity is only of the order $\mathcal{O}(m^2)$ which, in terms of $M$, becomes $\mathcal{O}(\log^2 M )$. In  \cite{babbush2018low}, a Fermion FFT algorithm is developed with reduced complexity of $\mathcal{O}(m)$. Furthermore, the initial wave function in \eqref{eq: schr} can be prepared on a quantum computer, by first computing its integral intervals in some subintervals.  This has been proposed in \cite{zalka1998efficient,wiesner1996simulations}. Ward et al. \cite{ward2009preparation} have shown how such computations can be done on a quantum computer. This makes quantum algorithms particularly appealing for solving \eqref{eq: schr}. In this section, we will outline the computational procedure.

\subsection{Time integration and  gate counts} 
For the time evolution, one can follow the operator splitting method \eqref{eq: splitting}, which in the context of quantum algorithms, belongs to Trotter splitting methods. They have been extensively studied \cite{childs2021theory}. In the splitting method \eqref{eq: splitting}, the implementation of the 
quantum Fourier transform (QFT) is also standard.  Regarding the diagonal unitary operators, $ U_V$ in \eqref{eq: splitting}, and $ \exp( -i  \frac{\dt\hbar}{2} \mu_k^2 ) $ 
in \eqref{eq: kin}, we will assume that they require ${J}(m)$ gates.{  We mention several alternatives here,
\begin{itemize}
    \item Bullock and Markov  \cite{shende2006synthesis,bullock2004asymptotically} have shown that a diagonal unitary operator can be implemented using $J=2^{m+1}-3=\CO{M}$ gates (see also \cite{schuch2003programmable}). This can be viewed as a full implementation, in the sense that the operator is implemented without approximations. 
   
    \item Diagonal unitary operators are sparse. Low and Wiebe \cite{low2018hamiltonian} presented a block-encoding algorithm for general Hamiltonian simulations,  which approximates the unitary operator with error within $\veps$. In \cite[Theorem 3]{low2018hamiltonian}, if we consider a diagonal time-independent Hamiltonian within only one time step, and choose $\veps$ to be comparable to $\dt^2$, then the gate complexity becomes $J=\CO{m \log \veps /\log\log\veps} $.  The method requires queries to an oracle to access the matrix elements. Accessing the oracle allows one to obtain values of $V(x).$ 
    
    \item Kassal et al. \cite{kassal2008polynomial} proposed to shift the potential so that it can be represented by integers between 0 and $2^{m_1}-1$ for some integer $m_1$. Then the diagonal operator can be implemented using an ancilla register with $m_1$ qubits, together with the quantum Fourier transform. Assume that $m_1= \CO{m}$, then $J(m)= \CO{m}$ as well. This work also took into account the complexity in computing the many-body potential, which leads to additional complexity that has a polynomial scaling in terms of $m$.
\end{itemize}
To leave the various available methods open, we just denote the gate count as $J(m).$ {Since the current approach already pursues approximations of wave functions, it is  likely that an exact implementation of the diagonal unitary operators is not necessary.   Rather, an approximate implementation with $J(m)=\CO{m}$ or $J(m)=\CO{\mathrm{poly}(m)}$ without affecting the overall accuracy is the desired approach. 
}  }

\medskip 
The number of gates required to obtain a solution of \eqref{eq: schr} with accuracy within $\veps$ hinges on an error estimate. 
The most widely used classical algorithm to approximate the time evolution of \eqref{eq: schr} is the Trotter splitting method.  Toward this end, one can split the Hamiltonian operator as follows, 
\begin{equation}\label{eq; AB}
  H= A + B, \quad   A:= -\frac{\hbar \nabla^2}{2}, \quad B:=  \frac{V}{\hbar},  
\end{equation}
and for each step, the solution represented by the operator $\ef{-i\dt H}$ is approximated by successively applying the following $2s$ operators,
\begin{equation}\label{eq: splitting}
    \ket{\psi^{n+1}} =U_{V}(c_1 \dt)U_{K}(d_1 \dt)
    U_{V}(c_2 \dt)U_{K}(d_2 \dt) \cdots 
    U_{V}(c_s \dt)U_{K}(d_s \dt)
    \ket{\psi^n}. 
\end{equation}
Here the coefficients $(c_1,c_2, \cdots, c_s)$ and $(d_1,d_2, \cdots, d_s)$ need to be selected based on certain order conditions.

The operators $A$ and $B$ involve the small parameter $\hbar$. To see how this plays a role in the Trotter error, one can consider the 1-step splitting, given by,
\begin{equation}\label{eq: split1}
    \ef{-itH} \approx \ef{-itA} \ef{-itB}. 
\end{equation}

A direct application of the BCH formula yields $\exp(-itA)\exp(-itB) = \exp(Z),$  where $Z$ can be expanded as 
$$Z = -it(A + B)  - \frac{t^2}{2}[A,B] + \frac{it^3}{12}([A,[A,B]]+[B,[B,A]]) + \frac{t^4}{24} [A,[B,[B,A]]] + \cdots.$$
Hence, the error is reflected in the commutator terms. At first sight, one finds from \eqref{eq; AB} that,
\[ 
[A,B]= \mathcal{O}(1),\;\; [A,[A,B]]= \mathcal{O}(\hbar),\;\; [B,[B,A]]= \mathcal{O}(\hbar^{-1}),  \;\;[A,[B,[B,A]]]= \mathcal{O}(1), \cdots.
\]
Clearly, these operators are of different magnitude. But this is further complicated by the fact that these operators have to be applied to the wave function, whose derivatives are not uniformly bounded with respect to $\hbar.$ The derivative may give rise  to negative powers of $\hbar.$ To incorporate the dependence on $\hbar$, Lasser and Lubich \cite{lasser2020computing} introduced an $\hbar$-dependent Sobolev norm. For each integer $q\ge 0$, this is defined as, 
\begin{equation}
    \norm{ \ket{\psi} }_{H_\hbar^q }^2 := \sum_{\alpha=0}^q  \hbar^\alpha  \left\| \px^\alpha \ket{\psi} \right\|^2.
\end{equation}
In addition, under the assumption that $V \in C^{2q}[0,L],$ they proved that \cite[Lemma 7.2]{lasser2020computing},
\begin{equation}\label{deriv-bd}
      \left\| \ket{\psi(\cdot,t)}  \right\|_{H_\hbar^q } \leq  (1+Ct)^q  \left\| \ket{\psi(\cdot,0)}  \right\|_{H_\hbar^q },
\end{equation}
for some constant $C$ that is independent of $\hbar$ and $t.$
In particular, this a priori estimate provides a bound on the derivatives of the wave function. Subsequently, in expressing the leading error terms, we will simply write $$\ket{ \px^{k} \psi}= \CO{\hbar^{-k}}.$$ For example, for an initial condition in the WKB form \eqref{wkb}, the corresponding solution at a later time will be in the same form, or a linear superposition of WKB forms. Therefore the derivatives follow the bound \eqref{deriv-bd}. Meanwhile, when the initial condition is smooth, the solution of \eqref{eq: schr} will remain smooth until caustics emerges, at which point, the derivative will be inversely proportional to $\hbar$ again.

The following estimate has been proved in  \cite[Theorem 4.1]{bao2002time}. 
\begin{theorem}
Assume  $V \in C^{2q}[0,L].$  Suppose that
 $\dx/\hbar =\mathcal{O}(1) $ and $\dt/\hbar =\mathcal{O}(1) $. For each integer $\ell < q$,  the error associated with \eqref{eq: split1} is bounded by,
\begin{equation}\label{eq: eSP1}
 \norm{ \ket{\psi^n} - \ket{\psi(\cdot,t_n)} } \le C_\ell L^{\frac12} n  \left( \left(\frac{\dx}{L \hbar }\right)^{\ell} +  
 \frac{\dt^2}\hbar\right). 
\end{equation}
\end{theorem} 
Here $C_\ell$ is proportional to $L^{-\frac12} \| \psi \|_{H_\hbar^\ell}$, and it is considered as $\CO{1}$. We included $L$ in the estimate so that the dependence on the domain size is reflected.  The first error term in \eqref{eq: eSP1} is the result from an interpolation error, since the finite Fourier transform corresponds to a trigonometric interpolation, while the second term represents the operator splitting error associated with \eqref{eq: split1}.

Let us first discuss the simplest case when $\ell=1.$ Given the initial wave function, our goal is to simulate the dynamics \eqref{eq: schr} until time $t$: $t=n\dt.$ 
It has been suggested in \cite{bao2002time} to implement the following meshing strategy:
\begin{equation}\label{eq: dx-dt}
 \frac{\dt}{\hbar} = \mathcal{O}\left( \frac{\veps}{L^{\frac12}t} \right), \quad \frac{\dx}{\hbar} = \mathcal{O}\left( \frac{L^{\frac12} \veps \dt}{t} \right),
\end{equation}
which is obtained by forcing both error terms to be of order $\veps$. In particular, with this choice of $\dt$, the number of steps is estimated to be,
\[
n = \frac{t}{\dt} = \CO{ \frac{L^{\frac12} t^2}{\hbar \veps}}. 
\]

 Based on these estimates, we can deduce the following gate complexity: 
\begin{theorem}
Given the error tolerance  $\veps$, the Schr\"odinger equation \eqref{eq: schr}, can be simulated using $m$ qubits  with
gate complexity $N_{Gates}$, given respectively by, 
\begin{equation}\label{eq: gateSP1}
m= \mathcal{O}\left(\log \frac{Lt^2}{\veps^2 \hbar^2}\right), \quad 
 N_{Gates}= \mathcal{O}\left( 2J(m)\frac{L^{\frac12} t^2}{\hbar \veps}   \right).
\end{equation} 
\end{theorem}

{ In this estimate, we first choose $\dt$ from \eqref{eq: dx-dt}, which subsequently determines how the grid size $\dx$ is selected. The selection of  $\dx$, in terms of $\veps, \hbar$ and $t$,  gives an estimate of the number of qubits required. In the gate count, the $2$ comes from the use of two unitary operators in the splitting. 

As an example, when the method \cite{bullock2004asymptotically} is used to implement the diagonal unitary operators, where $J(m)=\CO{M}$,  the total gate number becomes,
\begin{equation}\label{gates}
N_{Gates}= \mathcal{O}\left( \frac{L^{\frac32}t^4}{\veps^3\hbar^3}  \right).
\end{equation}
This is certainly unfavorable due to how it scales with $\hbar$ and $\veps,$ even for a one-dimensional problem. 

}

\smallskip

The leading error in \eqref{eq: eSP1} comes primarily from the commutator term,
\[ [A,B] \psi = \frac{1}2 \psi \partial_x^2 V + \partial_x V \partial_x\psi = \CO{\frac{1}{\hbar}}.    \] 
This can be improved by Strang splitting \eqref{eq: splitting-sp}  \cite{bao2002time}, which is a 2-step splitting method.
A direct application of the BCH formula yields, $\ef{-itB/2} \ef{-itA} \ef{-itB/2} = \ef{Z}$, with the operator $Z$ given by,
\[Z= -it(A+B) + \frac{it^3}{12}[B,[B,A]] - \frac{it^3}{24}[A,[A,B]] + \cdots. \]  

By direct substitutions of \eqref{eq; AB} into the commutatorslengthy calculations, we find that, 
\begin{equation}\label{V-smooth}
 \begin{aligned}
\; [B,[B,A]]\psi = &\hbar^{-1} (\px V)^2 \psi= \CO{\hbar^{-1} },\\
[A,[A,B]] \psi =& -\hbar( \frac{1}4  \psi \px^4 V  + \px^3 V \px \psi + \px^2\psi \px^2V) = \CO{\hbar^{-1}}.
\end{aligned}
\end{equation}

This analysis suggests that the time discretization error is of order $\frac{\dt^3}{\hbar}$. The extension of  \eqref{eq: eSP1} to Strang's splitting scheme \eqref{eq: splitting-sp}, which was proved in \cite[Theorem 7.5]{lasser2020computing}, is as follows,
\begin{equation}\label{eq: eSP2}
 \left\| \ket{\psi^n} - \ket{\psi(\cdot,t_n)} \right\| \le C_\ell  L^{\frac12} n \left(   \left(\frac{\dx}{\hbar L}\right)^
 \ell +  \frac{ \dt^3}\hbar\right). 
\end{equation}

If one considers the case $\ell=1$,  this error bound suggests the following strategy, 
\begin{equation}\label{eq: dx-dt-SP2}
 \frac{\dt^2}{\hbar} = \mathcal{O}\left( \frac{\veps}{L^{\frac12}t} \right), \quad \frac{\dx}{\hbar} = \mathcal{O}\left(  \frac{L^{\frac12} \veps \dt}{ t} \right),
\end{equation}
These choices suggest that one  picks $n= L^{\frac14} t^{\frac32}/(\hbar \veps)^{\frac12}.$

\begin{theorem}
Under the same assumptions, further assume that  estimate \eqref{eq: eSP2} holds.
Given the error tolerance  $\veps$, the Schr\"odinger equation \eqref{eq: schr} can be simulated by \eqref{eq: splitting-sp}  using $m$ qubits and gate complexity $N_{Gates}$, given respectively by,  
\begin{equation}\label{eq: gateSP2}
 m=\mathcal{O}\left(\log \frac{L^{\frac34}t^{\frac32}}{\veps^{\frac32} \hbar^{\frac32}}\right), \quad 
  N_{Gates}= \mathcal{O}\left(2 J(m) \frac{L^{\frac14} t^{\frac32}}{\hbar^{\frac12} \veps^{\frac12}} \right).
\end{equation} 
\end{theorem} 
We have included the factor $2$ owing to the fact that in the Strang splitting, the last operator at east step can be merged with the first operator in next step, thus reducing $3n$ unitary operators to $2n+1$.  Again, an exact implementation of the diagonal unitary operators leads to the following estimate,
\[ N_{Gates}= \mathcal{O}\left( \frac{2L t^3}{\veps^2\hbar^2}  \right). \]
Notice that compared to \eqref{gates}, this is an improved implementation in terms of the gate complexity because it is proportional to $(\veps \hbar)^{-2}$ (rather than $(\veps\hbar)^{-3}$). However, this is still an unfavorable scaling, and it is important to implement the diagonal unitary operators with $\CO{\mathrm{poly}(m)}$ complexity.

\smallskip

\begin{remark}
Our main emphasis has been placed on the role of $\hbar$ in the computational complexity.  It is also possible for the smoothness of the potential  $V$ to play a role. For instance, in light of the Trotter error associated with the Strang splitting and the commutators in \cref{V-smooth}, we see that the error bounds \eqref{eq: eSP2} carries a prefactor $V_{\text{max}}$,  given by,
 \begin{equation}
     V_{\text{max}} =\max \left\{ \norm{\partial_x V}_\infty^2, \norm{\partial_x^2 V}_\infty   \right\}.
 \end{equation}
 Therefore, the tolerance $\veps$ in the gate complexity has to be changed to $\veps/V_{\text{max}}$ accordingly. Similarly for the estimate \eqref{eq: eSP2}, the corresponding prefactor is $ V_{\text{max}} = \norm{\partial_x V}_\infty.$
\end{remark}

\begin{remark}
In addition to the gate complexity, it is also of practical interest to estimate the query complexity. The Strang spliting \eqref{eq: splitting-sp} consists of consecutively applied unitary diagonal matrices. The query complexity of  a diagonal unitary matrix depends on the implementation method. For example, suppose that we work with the oracle $O_D: \ket{j}\ket{0} \to \ket{j}\ket{D_{jj}},$ where $D_{jj}$ is a binary representation of the diagonal element, e.g., see \cite[Section II. A]{tong2021fast} and \cite{kassal2008polynomial}. Then by applying a phase gate, we obtain  $e^{i\dt D_{jj}} \ket{j} \ket{D_{jj}}$. Therefore the overall query complexity scales linearly with respect to the number of steps $n,$ and  according to \eqref{eq: dx-dt-SP2}, the query complexity is $\CO{L^{\frac14} t^{\frac32}/(\hbar \veps)^{\frac12}}$. Compared to the query complexity $\CO{\tau \log (\tau/\veps) /\log \log (\tau/\veps) }$ (with $\tau:= \max|D_{ii}| t$) \cite{berry2017exponential}, it is less efficient with respect to $\veps$, but more efficient with respect to $\hbar,$ since the operator $B$ in \eqref{eq; AB} is $\CO{\hbar^{-1}}$.
\end{remark}

\smallskip

For high order methods, Descombes et al. \cite[Theorem 1]{descombes2010exact}  derived order conditions for the time integration using operator-splitting scheme \eqref{eq: splitting}.  In particular, the splitting scheme is said to be $p$th order  if the local error 
is of the order $\CO{\frac{\dt^{p+1}}{\hbar}}$. In such high order methods, the leading error terms inevitably involve high order commutators. As an example, we consider the following commutators. 
\begin{prop}\label{prop}
The $(k+1)$st order commutator term $[A,[A,\cdots,[A, B]\cdots ],$ when applied to the wave function,  consists of the following terms,
\begin{equation}\label{eq: AAAB}
    [\underbrace{A,[A,\cdots,[A}_{k}, B]\cdots ] \ket{\psi}= \hbar^{k-1}\sum_{\overset{0\le   m_2 \leq k}{m_1+m_2 =2k} } c_{m_1, m_2} \px^{m_1} V \px^{m_2} \ket{\psi}. 
\end{equation}
In particular, it includes the term $-\px^{k} V \px^{k} \ket{\psi}$.
\end{prop}

This Proposition can be proved by induction. The case when $k=2$ has been proved in \cite{lasser2020computing}. Similarly, one can study the patterns from other commutators. It suggests that the local error associated with such terms is 
$\CO{\frac{\dt^{k+1}}{\hbar}}.$

Assuming that one has an $p$th order method with error bound, 
\begin{equation}\label{eq: eSP}
 \left\| \ket{\psi^n} - \ket{\psi(\cdot,t_n)} \right\| \le C_\ell  L^{\frac12} n \left(   \left(\frac{\dx}{\hbar L}\right)^
 \ell +  \frac{ \dt^{p+1}}\hbar\right), 
\end{equation}
then, the gate count becomes,
\begin{equation}\label{eq: general}
m=\mathcal{O}\left(\log \frac{L^{\frac12+\frac{1}{2p}}t^{1+\frac{1}p}}{\veps^{1+\frac1p} \hbar^{1+\frac1p}}\right), \quad 
  N_{Gates}= \mathcal{O}\left( 2p J(m) \frac{L^{\frac1{2p}} t^{1+\frac1p}}{\veps^{\frac1p} \hbar^{\frac1p}}  \right).
\end{equation} 
Here we set $\ell=1$ again, and  we assumed that the number of steps used in the splitting $s = \CO{p}$. 

When the diagonal unitary gates are implemented with $\CO{m}$ gates, one expects that by using a higher order splitting scheme, one can make the gate numbers  asymptotically approaches $ N_{Gates}= \mathcal{O}\left( 2p t \log \frac{L^{\frac12} t}{\veps \hbar}  \right).$ 

Finally, we consider the effect of the smoothness $\ell>1$. 
By setting
\begin{equation}\label{eq: dx-dt-SPpl}
 \frac{\dt^p}{\hbar} = \mathcal{O}\left( \frac{\veps}{L^{\frac12}t} \right), \quad \frac{\dx}{\hbar} = \mathcal{O}\left( \frac{L^{1-\frac{1}{2\ell}} \veps^{1/\ell} \dt^{1/\ell}}{t^{1/\ell}} \right),
\end{equation}
the corresponding estimates become,
\begin{equation}\label{eq: general-pl}
m=\mathcal{O}\left(  \log\frac{L^{ \frac{1}{2\ell} (1+\frac1p) } t^{ \frac1\ell ( {1+\frac1p})}}{\veps^{\frac1\ell(1+\frac1p)} \hbar^{1+\frac1p\frac1\ell}}  \right), \quad    
N_{Gates}= \mathcal{O}\left( 2p J(m) \frac{L^{\frac1{2p}} t^{1+\frac1p}}{\veps^{\frac1p} \hbar^{\frac1p}}  \right).
\end{equation} 
The query complexity is $\CO{p\frac{L^{\frac1{2p}} t^{1+\frac1p}}{\veps^{\frac1p} \hbar^{\frac1p}} }.$

The effect of this  on the number of gates is through $J(m)$:  When solutions are sufficiently smooth, the required number of qubits is much less, which then leads to a lower gate count. 

As examples of the operator splitting schemes \eqref{eq: splitting}, we list a few choices for each value of $s,$ for $1\le s \le 4$, in Table \ref{tab:split}. The three-step ($s=3$) method is obtained by choosing the coefficients to cancel the second and third order commutator terms, along with the term $[A,B,A,B]$ in the error expansion. The four-step method is the well known Yoshida's method from extrapolation \cite{yoshida1990construction}.

\begin{table}[thp]
\caption{Some examples of splitting schemes. }
\label{tab:split}
 \begin{tabularx}{\textwidth}{cll}
 \hline\hline
$s$ & Coefficients  & Leading error terms \\ 
 \hline
1 & {$c_1=d_1=1$} & \(\frac{\dt^2}{\hbar} [A,B]\) \\
\hline
2 &  {\( c_1=c_2=\frac12, d_1=1, d_2=0. \) }& \( \frac{\dt^3}{\hbar} [A,[A,B]],\; \frac{\dt^3}{\hbar} [B,[B,A]]\) \\
 \hline
3 & { \(
\begin{aligned}
  &  c_1= 0.26833, c_2=0.9197, c_3=1-c_1-c_2, \\
   &  d_1=0.63506, d_2= -0.1880, d_3=1-d_1 - d_2.
\end{aligned} \)}
  & \(\frac{\dt^4}{\hbar} [A,[A,[A,B]]], \; \frac{\dt^4}{\hbar} [B,[B,[B,A]]]\) \\
\hline
4 & { \(
\begin{aligned}
  &  c_1=c_4=\frac{1}{2(2-\sqrt[3]{2})},\; c_2=c_3=\frac{1}2 - c_1, \\
   & d_1=d_3= \frac{1}{2-\sqrt[3]{2}},\; d_4=0.
\end{aligned}\) }
 &  \( 
 \begin{aligned}
 \frac{\dt^5}{\hbar} [A,[A,[A,[A,B]]]],\; \frac{\dt^5}{\hbar} [A,[A,[B,[A,B]]]], \\ \frac{\dt^5}{\hbar} [B,[A,[B,[A,B]]]],\; \frac{\dt^5}{\hbar} [B,[B,[B,[B,A]]]].
 \end{aligned}
   \) \\
  \hline\hline
 %\end{tabular}
 \end{tabularx}
 \end{table}

\bigskip
\subsection{Gate counts for the computation of physical observables}
A remarkable observation in  \cite{bao2002time}  is that, if one is just interested in capturing correct physical observables, the step size can be chosen {\it independent} of $\hbar$. This observation was interpreted by using the Wigner transformation approach and making a connection to the Liouville equation for classical Hamiltonian dynamics. 
This issue has been later further investigated in \cite{golse2021convergence,lasser2020computing} mathematically rigorously. For instance, for the Strang splitting method \eqref{eq: splitting-sp}, combined with the  pseudo-spectral method, the following error bound has been obtained \cite[Theorem 7.4]{lasser2020computing},
\begin{equation}\label{eq: obs1}
     \left\| \langle A \rangle_{\ket{\psi^n}} - \langle A \rangle_{\ket{\psi(\cdot,t_n)}} \right\| \le C L^{1/2} n \left(  \left( \frac{\dx}{\hbar L} \right)^\ell  +  \dt^3 + \dt \hbar^2 \right). 
\end{equation}

Notice that the analysis in \cite{golse2021convergence,lasser2020computing}  only considered the continuous case, i.e., without the pseudo-spectral method in space. Therefore, we included the interpolation error term to account for the spatial discretization error.

In this case, we have,
\begin{theorem}\label{obs}
Given the error tolerance  $\veps$, assume that $\hbar < \sqrt{\veps/t}$. Then, observables from the Schr\"odinger equation \eqref{eq: schr}, can be simulated using 
\begin{equation}
    m= \mathcal{O}\left( \log \frac{L^{\frac3{4\ell}} t^{\frac3{2\ell}}} {\veps^{\frac{3}{2\ell}} \hbar} \right),
\end{equation}
 qubits, with
gate complexity,
\begin{equation}
  N_{Gates}= \mathcal{O}\left( 2J(m)   \frac{L^{\frac14}t^{\frac32}}{\veps^{\frac12}} \right),
\end{equation} 
and query complexity of $\CO{\frac{L^{1/4}t^{3/2}}{\veps^{1/2}}}$. 
\end{theorem}

In terms of the dependence on $\hbar^{-1}$, in the worst case scenario where the full implementation has to be used for the diagonal unitary operators, the gate complexity is $\CO{\frac1\hbar}$. On the other hand, when $J(m)= \CO{\text{poly}(m)}$, the dependence  is only polylog.  This is a considerable reduction compared to \eqref{eq: gateSP2}.

This reduction in the computational complexity can be attributed to the fact that in each step of the operator-splitting scheme, the quantum dynamics \eqref{eq: schr} is solved exactly \cite{golse2021convergence,lasser2020computing}. Other approximation methods, including traditional Runge-Kutta methods or  Crank-Nicolson scheme, do not have this property. For instance, the numerical results in \cite{markowich1999numerical} indicate that in the Crank-Nicolson's scheme, the step size $\dt$ has to be much smaller than $\hbar$ in order to have the correct limiting behavior. In contrast, as suggested by the error bound \eqref{eq: obs1}, $\dt$ in the operator-splitting scheme can be chosen independent of  $\hbar$ \cite{bao2002time}.

Another necessary step to obtain observables is to make measurements upon the completion of the quantum algorithm. Each observable corresponds to a Hermitian operator $A_j$, $1\leq j \leq M$, and we aim to estimate $ \bra{\psi(T)}A_j\ket{\psi(T)}$. In order to obtain an estimate within $\veps$ with high probability, a direct sampling
approach, based on the Markov inequality, would requires repeating the algorithm $\frac{M}{\veps^2}$ times. To estimate the observables more efficiently, we consider the approach by Huggins et al. \cite[Theorem 1]{HMW+21}.

\begin{theorem}[{\cite[Theorem 1]{HMW+21}}]
Let $\{A_j\}$ be a collection of $M$ Hermitian operators on $m$ qubits, with spectral norms $\norm{A_j} \leq 1$ for all $j$. There exists a quantum algorithm that, for any $N$-qubit quantum state $\ket{\psi}$ prepared by a quantum circuit $U_{\psi}$, outputs estimates $\langle\widetilde{A_j}\rangle$ such that $\lvert \langle\widetilde{A_j}\rangle - \bra{\psi}A_j\ket{\psi}\rvert \leq \epsilon$ for all $j$ with probability at least $2/3$, using $\mathcal{O}(\sqrt{M}/\epsilon)$ queries to $U_{\psi}$ and $U_{\psi}^{\dag}$, along with $\mathcal{O}(\sqrt{M}/\epsilon)$ gates.
\end{theorem}

\begin{coro}
Under the same conditions in \cref{obs}, for any $\delta \in (0,1) $, observables from the Schr\"odinger equation \eqref{eq: schr} can be obtained with probability at least $1-\delta$ from a quantum algorithm with gate complexity \begin{equation}
  N_{Gates}= \mathcal{O}\left( 2J(m)   \frac{\sqrt{M} L^{\frac14}t^{\frac32}}{\veps^{\frac32}} \log \frac{1}{\delta} \right).
\end{equation}

\end{coro}

\subsection{High-dimensional cases}

It is straightforward to extend the pseudo-spectral methods to  high-dimensional cases with dimension $d$ by using tensor products. More specifically, one can consider $M_q$ uniformly spaced grid points with $1\leq q \leq d$.    A direct implementation requires $M_1 M_2 \cdots M_d$ grid points, which depends exponentially on the dimension $d$. Fortunately,  the corresponding number of qubits  $m=\sum_{q=1}^d \log_2M_q,$ which grows only linearly.   

 A similar error bound for the trigonometric interpolation can be found in \cite[Theorem 5]{pasciak1980spectral}. Therefore, the error bounds from previous sections still hold. For simplicity, we can consider a $d$-dimensional cube as the domain and choose $M_q=M$.
 Then we can extend the bound \eqref{eq: eSP} to
 \begin{equation}\label{eq: eSPd}
 \left\| \ket{\psi^n} - \ket{\psi(\cdot,t_n)} \right\| \le C_\ell  L^{\frac{d}2} n \left(   \left(\frac{\dx}{\hbar L}\right)^
 \ell +  \frac{ \dt^{p+1}}\hbar\right). 
\end{equation}
Furthermore, we generalize \eqref{eq: obs1}  to,
 \begin{equation}\label{eq: obsd}
     \left\| \langle A \rangle_{\ket{\psi^n}} - \langle A \rangle_{\ket{\psi(\cdot,t_n)}} \right\| \le C L^{\frac{d}2}  n \left(  \left( \frac{\dx}{\hbar L} \right)^\ell  +  \dt^3 + \dt \hbar^2 \right). 
\end{equation}

In this case, we have
\begin{equation}\label{eq: nqd}
    m= \mathcal{O}\left( d \log \frac{L^{\frac{3d}{4\ell}} t^{\frac3{2\ell}}} {\veps^{\frac{3}{2\ell}} \hbar} \right),
\end{equation}
 qubits, with
gate complexity,
\begin{equation}\label{eq: ngd}
  N_{Gates}= \mathcal{O}\left( 2J(m)   \frac{L^{\frac{d}4}t^{\frac32}}{\veps^{\frac12}} \right).
\end{equation} 

In classical algorithms, the computational complexity, scales $n^d$ at each time step. The estimate \eqref{eq: obsd} suggests that  $n \geq \frac{1}{\hbar \veps^{1/\ell}}$, yielding a complexity of the order $\left(\frac{1}{\hbar \veps^{1/\ell}}\right)^d$. This exponential scaling is unfavorable when $h\ll 1$ or $\epsilon \ll 1.$ On the other hand, in the complexity from the quantum algorithm \eqref{eq: ngd} the exponent of $\epsilon$ does not change with the dimension $d$.

\section{Numerical results}

\subsection{Numerical tests on several Trotter splitting methods}
We first show results from the operator splitting methods on a classical computer. The objective is to demonstrate the solution behavior and the performance of the operator splitting methods in computing the wave functions and observables.  In the example, which is similar to the second example in \cite{bao2002time}, we solve \eqref{eq: schr} in [-2,2], with the initial condition written in a WKB form: $\ket{\psi_0}=A_0 \ef{S_0(x)/\hbar}$. We consider the following initial amplitude and phase functions,
\[ A_0(x) =  \exp(-25(x-0.5)^2 ), \quad S_0(x)= - \frac15 \ln \left(\exp (5 (x-0.5)) + \exp (-5 (x-0.5)) \right). \]
In addition, we choose the harmonic potential $V=\frac{x^2}2.$

In addition to the wave function, we will monitor the amplitude and the current as  observables. They are defined respectively as follows,
\begin{align}%\label{eq: den-cur}
n(x,t) =&  |\psi(x)|^2, \label{eq: den} \\ 
\label{eq: cur}J(x,t)=&    \hbar \text{Im} (\overline{\psi} \px \psi).  
\end{align}

\begin{figure}[htbp]
\begin{center}
\includegraphics[scale=0.19]{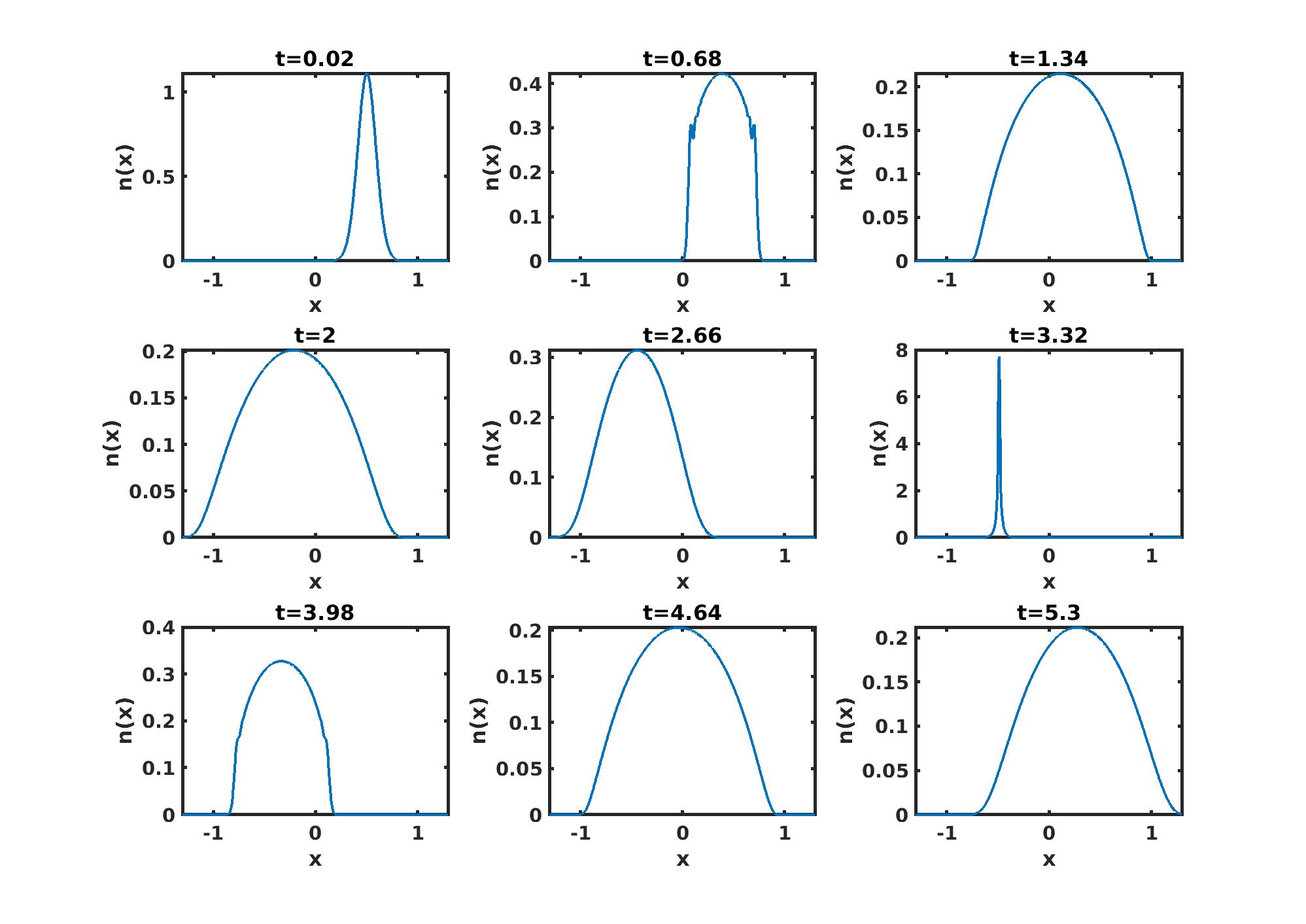}
\caption{The time evolution of the amplitude. }
\label{fig: ampH}
\end{center}
\end{figure}

We first show in Fig. \ref{fig: ampH} the history of the amplitude of the solution \eqref{eq: den}. Here $\hbar=0.003$. We used fine mesh {{$\Delta x=2^{-16}$}} and small step size $1/6400$ to fully resolve $\hbar$ so that the solution can be regarded as the `exact' solution. The dynamics involves a few  interesting stages. First the initial Gaussian profile collapses into a peak, which evolves into a dome shape, and then spreads out.  Around $t=2$, the amplitude begins to refocus, and it  develops a peak around $t=3.32.$ The peak then leads to an oscillatory pattern, which subsequently becomes more smooth.     

\begin{figure}[htbp]
\begin{center}
\includegraphics[scale=0.19]{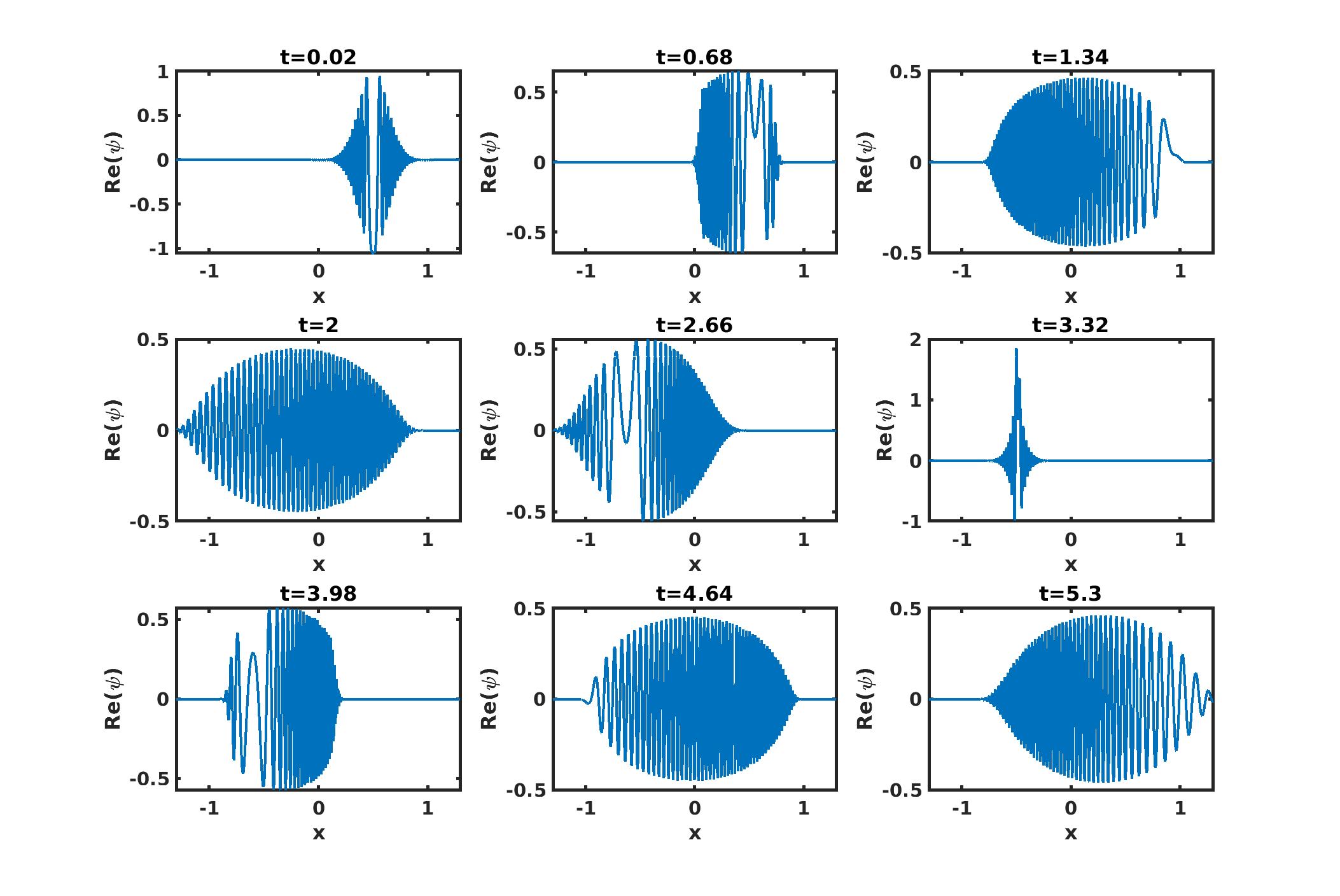}
\caption{The time evolution of the amplitude. }
\label{fig: realH}
\end{center}
\end{figure}

As a comparison, Fig. \ref{fig: realH} shows the history of the wave function (the real part), which exhibits much faster oscillations due to the small value of $\hbar.$

\begin{figure}[htbp]
\begin{center}
\includegraphics[scale=0.14]{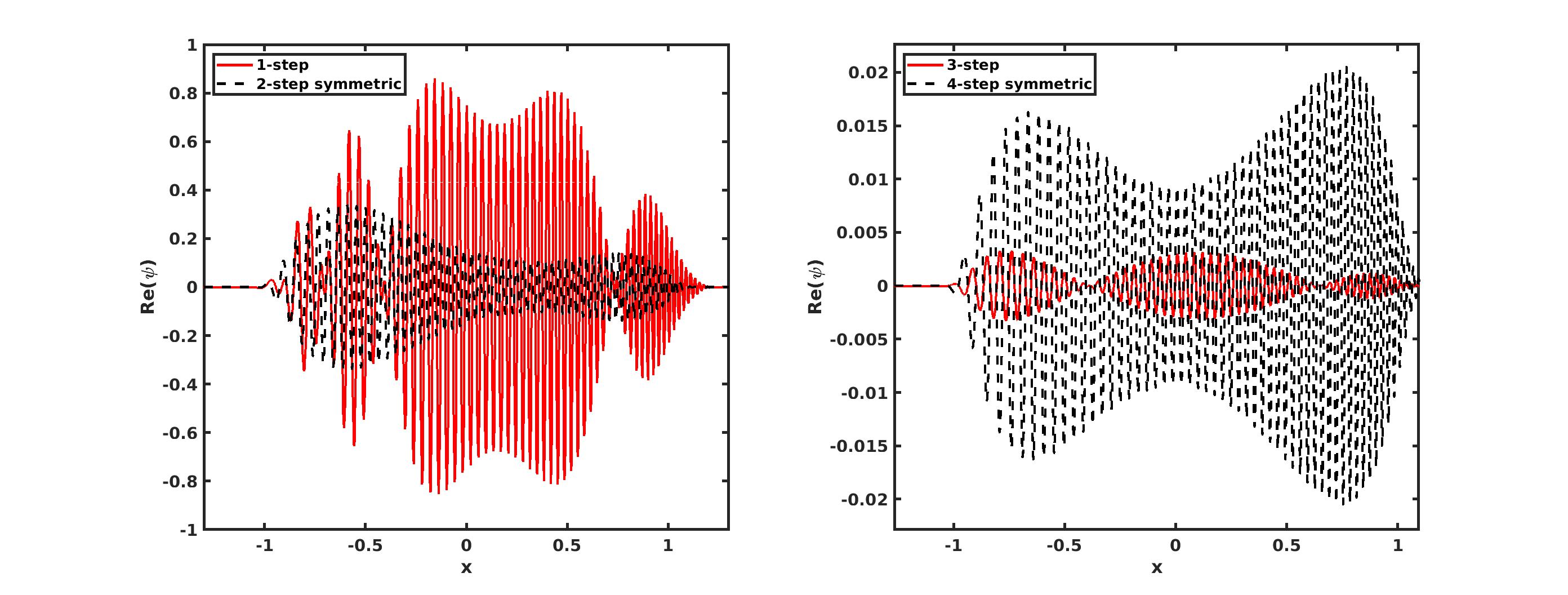}
\caption{The error of the real parts of the wave function. Results computed from the one and two-step splitting schemes (Left); The three-step non-symmetric scheme and the Neri's $4$th order scheme (right). }
\label{fig: realE}
\end{center}
\end{figure}

To test the error from the operator splitting methods, especially those  listed in Table \ref{tab:split}, we choose $\dx=2^{-9}$ and $\dt=0.2$. Notice that $\dx$ is less than $\hbar$, but {\it $\dt$ is much larger than $\hbar$}. Fig. \ref{fig: realE} shows the error in the real part of the wave function from these methods. One can see that the one-step and two-step methods produced significant error. The three- and four-step methods, due to the increased order, have reasonable accuracy. 

\begin{figure}[htbp]
\begin{center}
\includegraphics[scale=0.12]{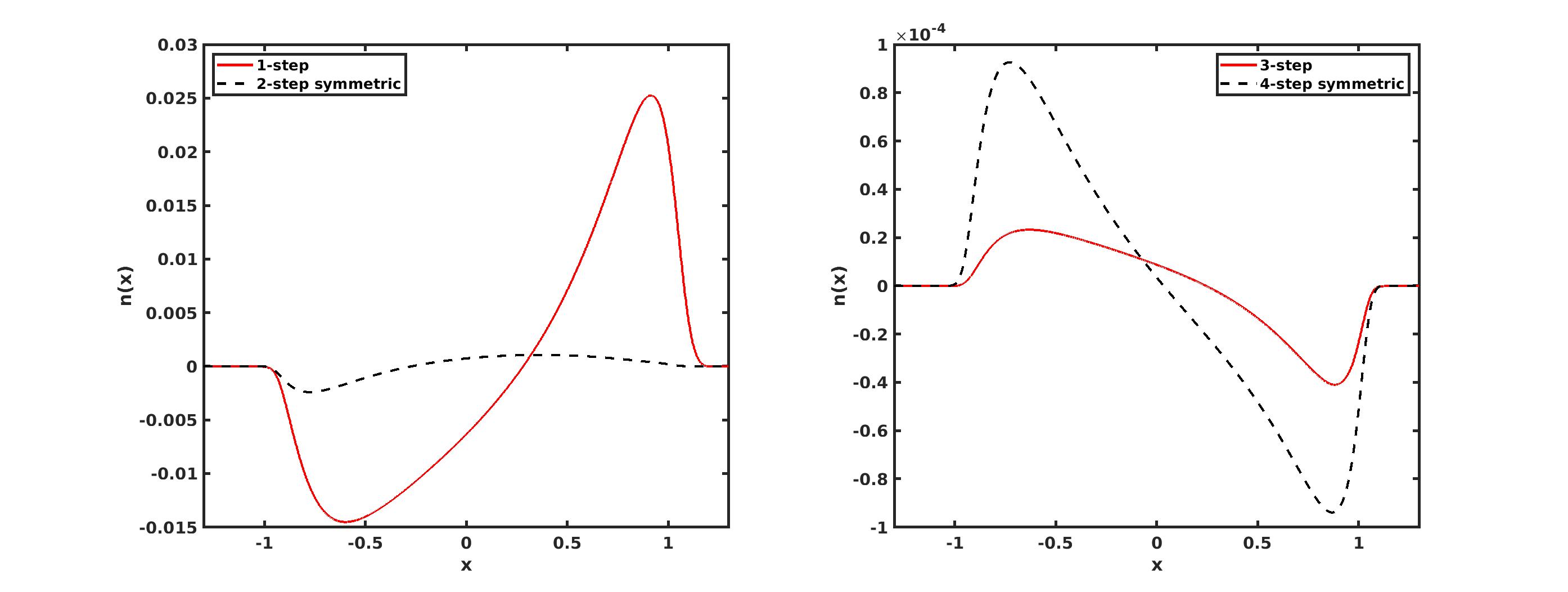}
\caption{The error of the amplitude $n(x,t)$.  Results computed from the one and two-step splitting schemes (Left); The three-step non-symmetric scheme and the Neri's $4$th order scheme (right).}
\label{fig: ampE}
\end{center}
\end{figure}

\begin{figure}[htbp]
\begin{center}
\includegraphics[scale=0.12]{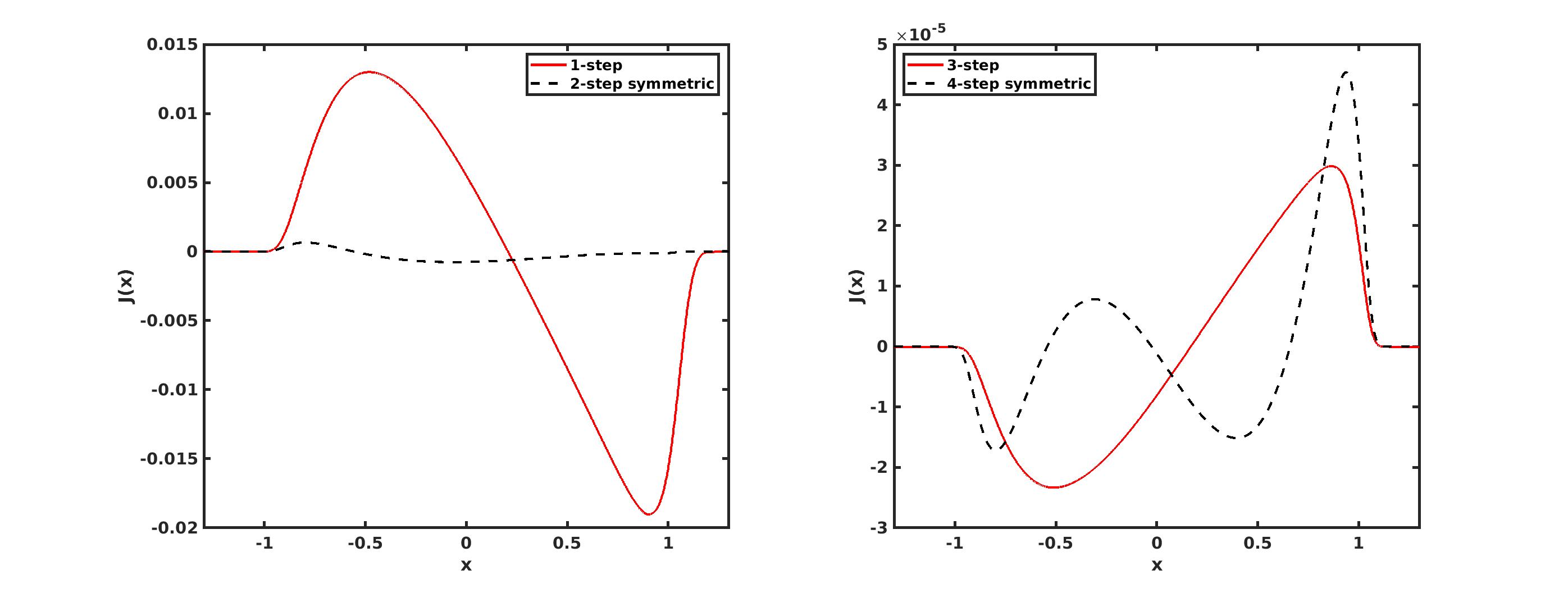}
\caption{The error of the current. Results computed from the one and two-step splitting schemes (Left); The three-step non-symmetric scheme and the Neri's $4$th order scheme (right). }
\label{fig: curE}
\end{center}
\end{figure}

Now we turn to the computed observables. Figs. \ref{fig: ampE} and \ref{fig: curE} show the errors in the amplitude and the current, respectively. We observe that overall, the error is much smaller than the error associated with the wave function. Surprisingly, the three-step non-symmetric method performs slightly better than the four-step symmetric method in computing the amplitude. Upon close inspection, we find that this  is due to the harmonic potential used in this test. In this special case, we have $[B,B,A]=(\partial_x V)^2 \ket{\psi}=x^2 \ket{\psi}.$ As a result, the commutator term $[B,B,B,A]$ in the error is zero. In addition, the terms in $[A,A,A,B]$, according to Proposition \ref{prop}, involves $-\partial_x^k V$ with $k\geq 3$, implying that the error term is also zero.

\subsection{Numerical experiments on quantum simulators}
We also implemented the operator splitting algorithm \eqref{eq: splitting} using the quantum simulator package Qiskit \cite{koch2019introduction,wille2019ibm}.

{ In order to measure observables, we need to express them as Hermitian operators. Using the Fourier transform, we can express the amplitude \eqref{eq: den} in the Fourier basis,
\begin{equation}
    |\braket{j}{\psi}|^2 = \frac{1}{M} \sum_{k=0}^{M-1}\sum_{k'=0}^{M-1} e^{i2\pi j(k-k')/M }\braket{\psi}{k'}  \braket{k}{\psi}. 
\end{equation}
Therefore, we can define the position operators as,
\begin{equation}
 n(x_j)= \bra{\psi}    \wh{n}_j \ket{\psi}, \; \wh{n}_j := \frac{1}{M}\sum_{k=0}^{M-1}\sum_{k'=0}^{M-1} e^{i2\pi j(k-k')/M } \ketbra{k'}{k}.
\end{equation}
The formula on the right hand side corresponds exactly to the matrix elements of the density operator  $\hat{n}_j$  in the Fourier basis.

Similarly, we can express the current \eqref{eq: cur} in the Fourier basis. 
\begin{equation}
 J(x_j)= \bra{\psi}    \wh{J}_j \ket{\psi}, \; \wh{J}_j := \frac{\hbar }{2Mi} \sum_{k=0}^{M-1}\sum_{k'=0}^{M-1} \big( e^{i2\pi j(k-k')/M }\mu_k  - e^{-i2\pi j(k-k')/M } \overline{\mu_{k'}} \big) \ketbra{k'}{k}.
\end{equation}
Here $\mu_k$ is from Eq. \eqref{eq: mu}.
}

\begin{figure}[htbp]
\begin{center}
\includegraphics[scale=0.14]{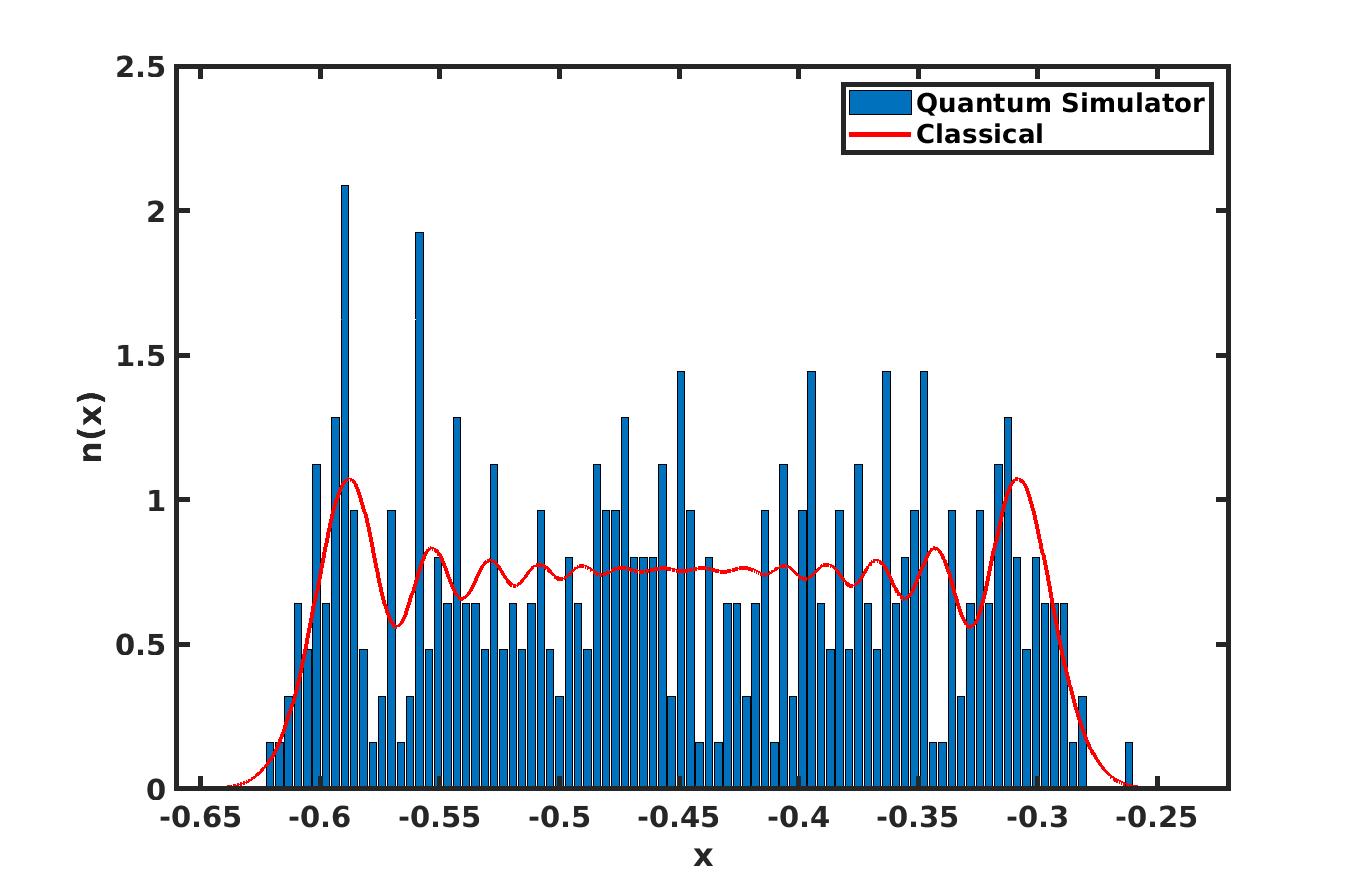}\\
\includegraphics[scale=0.14]{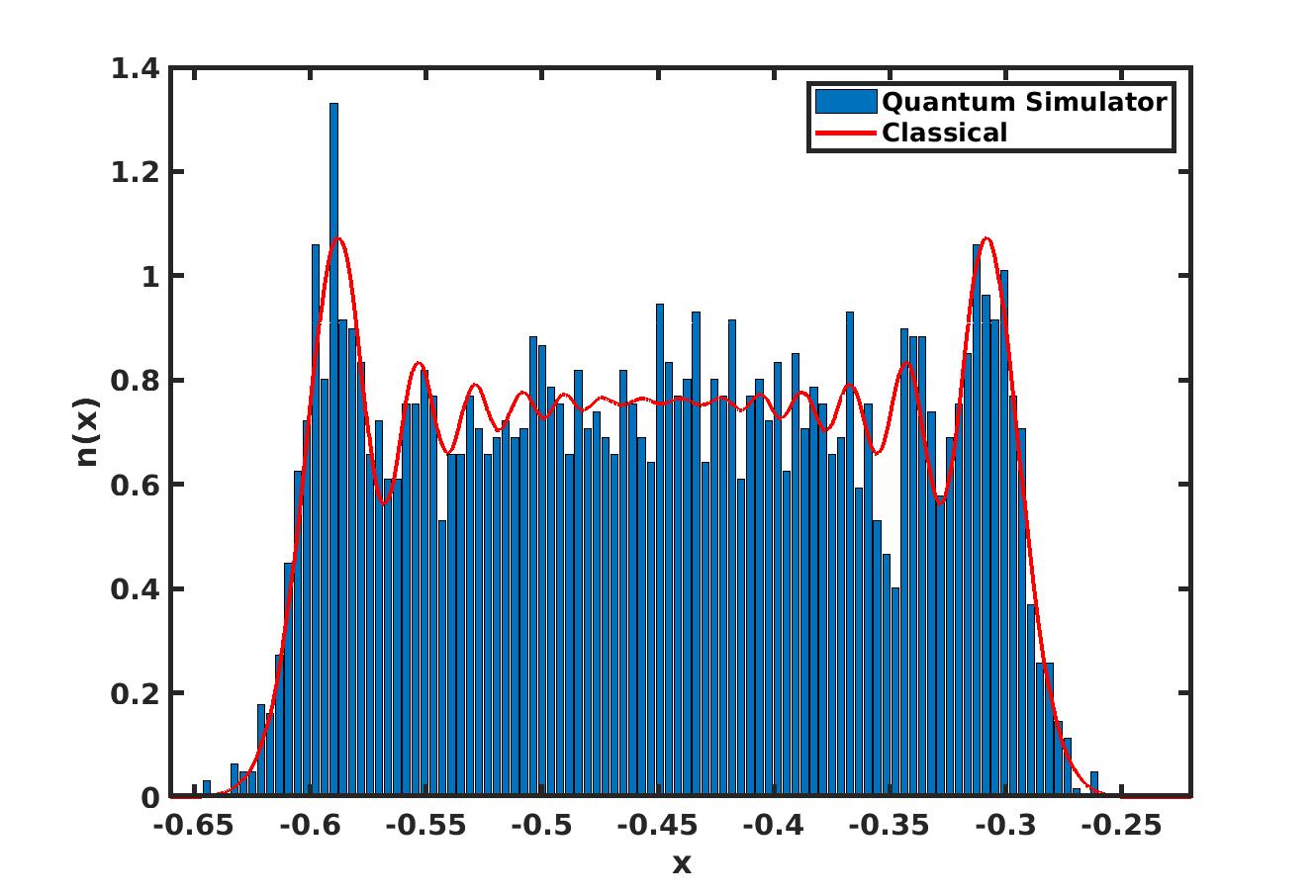}\\
\includegraphics[scale=0.14]{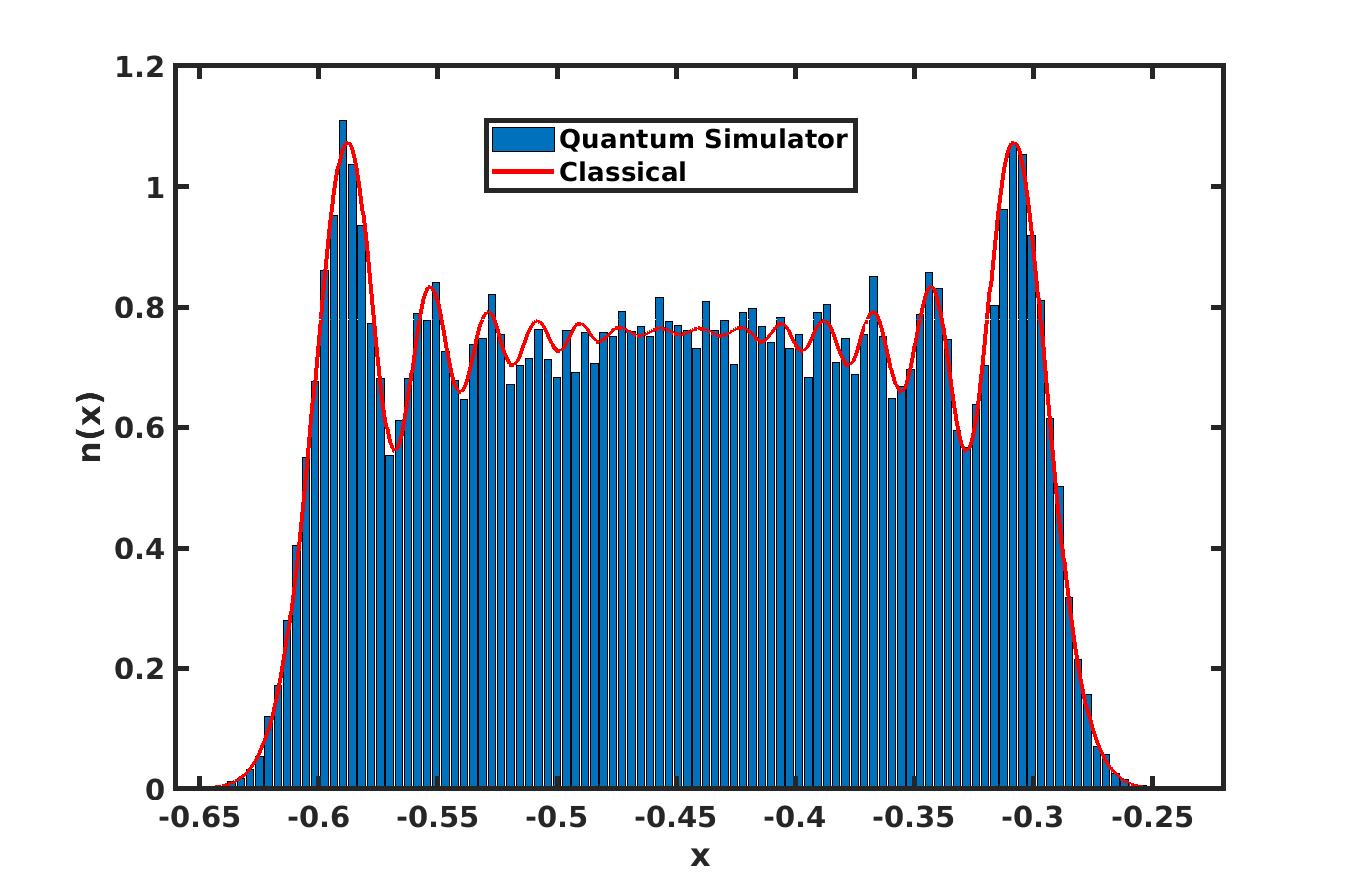}
\caption{The amplitude computed using the Strang splitting. From top to bottom: 400 shots, 4000 shots and 40000 shots. }
\label{fig: FIG2b}
\end{center}
\end{figure}

We consider the same test problem as in the previous section using Strang's splitting \eqref{eq: splitting-sp}. But we choose $\dx=2^{-8}$ by using 10 qubits. This allows us to better visualize the counts using bar plots from the measurement steps.  We run the dynamics using $72$ steps with step size $\dt=0.05.$   The result is also compared with the amplitude from the classical simulation with fine mesh and small step size.
Fig. \ref{fig: FIG2b} shows the amplitude from the measurement. Specifically, after the quantum simulation, we measure the magnitude of the wave function using multiple shots. The measurements yield the number of times each state in the computational basis ($\ket{j}$) appears. Therefore, we show the bar plot in the Figure.  In particular, we run the experiment with 400, 4000, and 40000 shots in the measurement step. Clearly, the number of shots have a strong impact on the computed amplitude.  With 40000 shots, the result agrees very well with the solution obtained with the classical algorithms. 

\subsection{Potential Applications}

One important application of the Schr\"odinger equation \eqref{eq: schr} in the semi-classical regime $\hbar\ll 1$ is the Born-Oppenheimer molecular dynamics. In this case, Eq. \eqref{eq: schr} can be derived from the Born-Oppenheimer approximation, reducing the problem to the nuclei dynamics governed by a Schr\"odinger equation similar to \eqref{eq: schr}. In this setting, $\hbar^2$ represents the ratio between the electron and nuclei mass. This has been the primary motivation for conducting the numerical analysis and computation of Eq. \eqref{eq: schr}. One application considered in \cite{sholl1998generalized,jin2011eulerian} is the scattering of a light particle off of a solid surface. We hypothesize a computation that starts with a Gaussian wave packets,
\[
\psi = \ef{ - \frac{\|x - x_0 \|^2 }{\gamma } - \frac{ik_0\cdot x}{\hbar}}.
\]

Here we chose  $\gamma=0.5$ \AA as in \cite{sholl1998generalized}. Based on this parameter, we introduce a three-dimensional computational domain of the size $[0,10]$\AA $\times [0,10]$ \AA $\times [0,10]$\AA. In addition, the nuclei motion is often on the time scale of femto seconds. We set $t=1000$ to compute the dynamics on the scale of pico seconds as done in \cite{sholl1998generalized}. To estimate the computational resources, we consider moderate values $\ell=p=2.$   For the parameter $\hbar$, we choose $\hbar=10^{-1}, 10^{-2}$ and $10^{-3}$, to cover a wide range of possible values. For instance, several molecules were considered in \cite{ishikawa2012accurate} to study the dynamics with and without the Born-Oppenheimer approximation. For these specific example, the correspond values of $\hbar$
range from 0.0135 ($HT^+$ molecule) to 0.0233 ($H_2^+$ molecule). 
 
\begin{figure}[htbp]
\begin{center}
\includegraphics[scale=0.24]{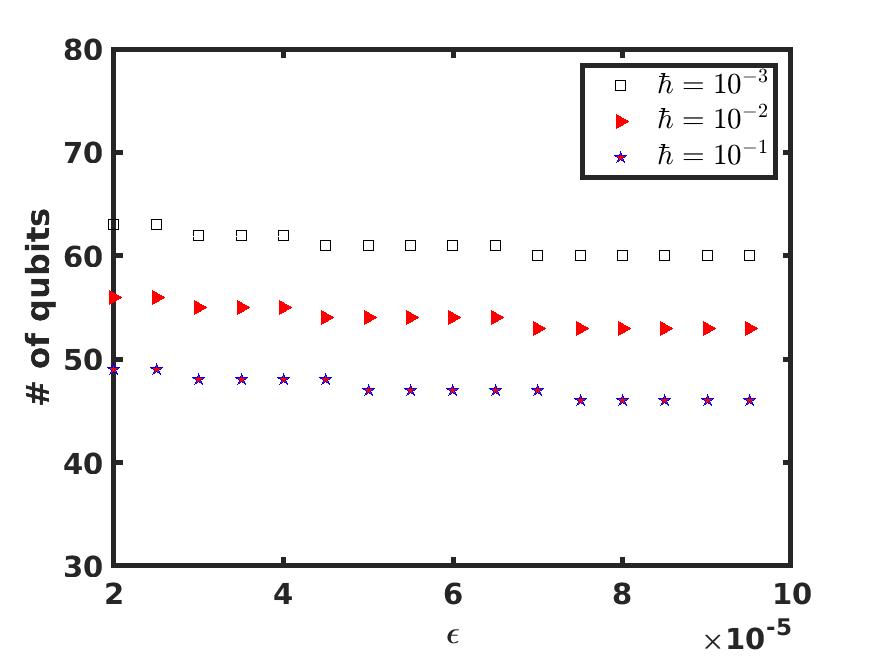}
\caption{ The estimated number of qubits for the scattering problem for various choices of $\veps$ and $\hbar.$ }
\label{fig: scattNq}
\end{center}
\end{figure}

Fig. \ref{fig: scattNq} depicts estimates of the number of qubits for various choices of the precision $\veps$. Here we assume that $J(m)=\CO{m},$ and used the bound \eqref{eq: obsd} and \eqref{eq: nqd}. In this case, the estimates are less sensitive to the precision $\veps$ than the parameter $\hbar.$ It is clear from these estimates  that such computation can be done with relatively small quantum computers.   Since the dependence on the dimension $d$ is linear, one can extend the computation to higher dimensions and use the algorithm to solve the many-body Schr\"odinger equations. 

\begin{figure}[htbp]
\begin{center}
\includegraphics[scale=0.24]{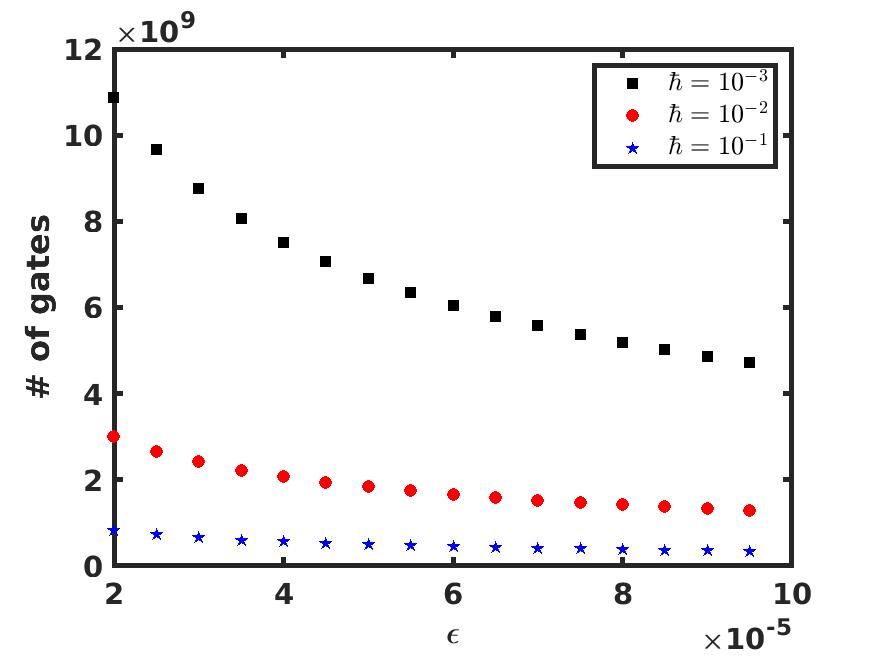}
\caption{ The estimated number of gates for the scattering problem to compute the wave function with error within  $\veps$. }
\label{fig: scattNgWave}
\end{center}
\end{figure}

\begin{figure}[htbp]
\begin{center}
\includegraphics[scale=0.24]{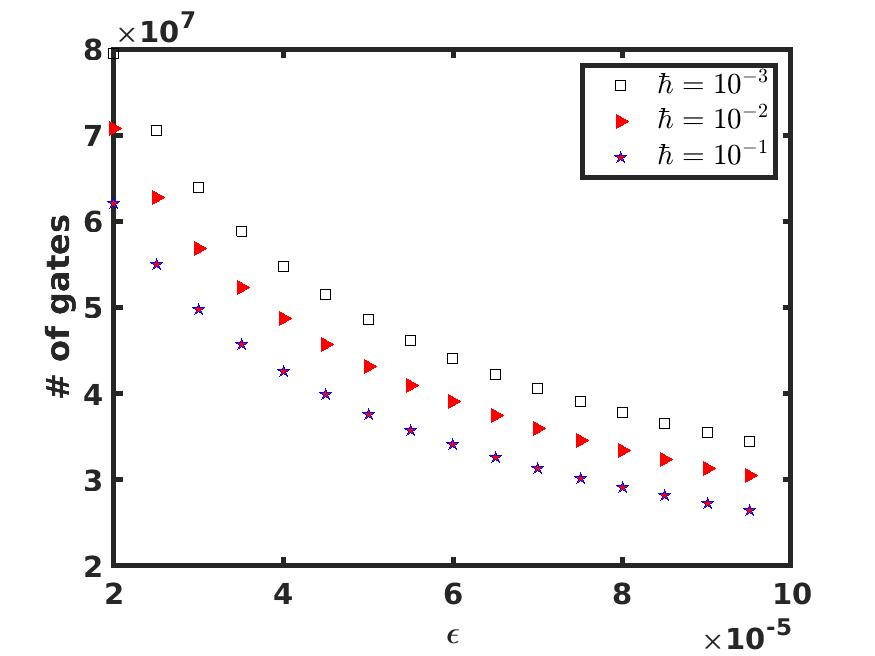}
\caption{ The estimated number of gates for the scattering problem to compute observables with error within  $\veps$. }
\label{fig: scattNg}
\end{center}
\end{figure}

Next we use the bound \eqref{eq: eSPd} to obtain estimates for the number of gates that are needed to compute the wave functions with accuracy $\veps.$  The estimates are shown in Fig. \ref{fig: scattNgWave}. Similarly, we use 
the bound \eqref{eq: ngd} to estimate the gate numbers for computing observables.
Fig. \ref{fig: scattNg} demonstrates these estimates. Clearly, computing observables requires resource that is two orders of magnitude less. 

\section{Summary and Discussions}

This work examined the Trotter splitting algorithms that alternate between the kinetic and potential energy terms in the Hamiltonian operator. In particular, the focus is on the semiclassical regime of the time-dependent Schro\"odinger equation, especially the dependence of the complexity on the small parameter $\hbar.$
{
The quantum advantage for this problem, compared to the classical counterpart,  comes from the ability to simulate the quantum state, even when $\hbar$ is very small. The number of qubits only scales logarithmically with respect to $\hbar$. Meanwhile, the efficiency of the quantum algorithm hinges critically on the ability to efficiently implement the diagonal unitary operators in the operator splitting schemes.

An alternative time integration method is the Crank–Nicolson method, which can be extended to high order Runge-Kutta Gaussian methods that preserve the norm of the wave function. Unlike the operator splitting methods \eqref{eq: splitting-sp}, these methods lead to linear system of equations that is solved at each step. However, in quantum algorithms, these linear systems can be handled by quantum linear solvers, such as the HHL method \cite{harrow2009quantum} and the block encoding approach \cite{childs2017quantum}. This approach has been pursued in \cite{childs2019quantum} for linear ODE systems. Although Crank–Nicolson type methods are not robust in the regime $\hbar$ is very small \cite{markowich1999numerical}, it would be of theoretical interest to  find the condition number of the corresponding linear system and the complexity of such methods in such a multiscale setting. }

This paper focused on Hamiltonian simulations in the semi-classical setting. In the case when $\hbar=\CO{1}$, the results will still apply. We are not aware of problems where $\hbar \gg 1$. But there are problems  where $\hbar$ or $m$ is variable \cite{petit1988interpretation}. Understanding the performance of quantum algorithms in those regimes would also be an interesting direction. 

\section*{Acknowledgement}
Jin's research was supported by NSFC grant No. 12031013,  Shanghai Municipal Science and
Technology Major Project 2021SHZDZX0102, and Innovation Program of Shanghai Municipal Education Commission 2021-2025. Li's research on quantum algorithms is supported by the National Science Foundation Grants DMS-2111221. Li would like to thank Dr. Chunhao Wang for fruitful discussions.  Liu's research is supported by the Shanghai Pujiang Talent Grant (no. 20PJ1408400) and the NSFC International Young Scientists Project (no. 12050410230). Liu is also supported by the Innovation Program of the Shanghai Municipal Education Commission (no. 2021-01-07-00-02-E00087), the Shanghai Municipal Science and Technology Major Project (2021SHZDZX0102) and the Natural Science Foundation of Shanghai grant 21ZR1431000.

\bibliographystyle{quantum}
\bibliography{schr,splitting}

\end{document}